\documentclass[reprint, superscriptaddress, amsmath,amssymb, aps, pra, floatfix]{revtex4-1}
\usepackage[T1]{fontenc} 
\usepackage{mathptmx}
\usepackage{graphicx}
\usepackage{dcolumn}
\usepackage{bm}
\usepackage[colorlinks, breaklinks, citecolor=blue,linkcolor=black,urlcolor=black]{hyperref}

\begin{document}

\title{Saturated absorption spectroscopy of buffer-gas-cooled Barium monofluoride molecules}
\author{Wenhao Bu}
\author{Yuhe Zhang}
\author{Qian Liang}
\author{Tao Chen}
\affiliation{%
Interdisciplinary Center of Quantum Information, State Key Laboratory of Modern Optical Instrumentation, and Zhejiang Province Key Laboratory of Quantum Technology and Device of Physics Department, Zhejiang University, Hangzhou 310027, China
}%
\author{Bo Yan}
\email{yanbohang@zju.edu.cn}
\affiliation{%
Interdisciplinary Center of Quantum Information, State Key Laboratory of Modern Optical Instrumentation, and Zhejiang Province Key Laboratory of Quantum Technology and Device of Physics Department, Zhejiang University, Hangzhou 310027, China
}%

\date{\today}

\begin{abstract}

We report an experimental investigation on the Doppler-free saturated absorption spectroscopy of buffer-gas-cooled Barium monofluoride (BaF) molecules in a  4~K  cryogenic cell. The obtained spectra with a resolution of 19~MHz, much smaller than previously observed in absorption spectroscopy, clearly resolve the hyperfine transitions. Moreover, we use these high-resolution spectra to fit the hyperfine splittings of excited $A(v=0)$ state and find the hyperfine splitting of the laser-cooling-relevant $A^2\Pi_{1/2}(v=0, J=1/2, +)$  state is about 18 MHz, much higher than the previous theoretically predicted value. This provides important missing information for laser cooling  of  BaF molecules.

\end{abstract}

\maketitle

\section{introduction}

Cold molecules, due to the additional internal degrees of freedom and the long-range dipolar interactions, provide novel potential applications and sophisticated understandings in controlled chemistry reactions \cite{Krems2008, Ospelkaus2010,Ni2019,Liu2021}, precision measurement \cite{Safronova2018, Cairncross2019, Cairncross2017, Altunta2018, Kozyryev2017, Lim2018,Vutha2018}, quantum simulation and computation \cite{Demille2002, Rabl2006, Andre2006}, and many-body physics \cite{Wang2006, Buechler2007}. In the last decade, various cooling techniques have been developed to achieve dense cold molecular samples, for example, buffer gas cooling \cite{Kantrowitz1951, Hutzler2012}, Stark and Zeeman deceleration \cite{Meerakker2012}, and assembling molecules from two ultracold atoms \cite{Ni2008}. Despite these efforts, producing cold molecule samples is limited to several kinds of species and remains an ongoing challenge.

Direct laser cooling as an alternative route, mainly for those molecules with a quasi-cycling transition \cite{DiRosa2004,Stuhl2008}, has been largely developed to extend the coolable molecular species. In recent years, the molecules, SrF \cite{Barry2014}, CaF \cite{Truppe2017, Anderegg2017} and YO \cite{Collopy2018}, have been successfully loaded into three-dimensional magneto-optical traps with temperatures at few millikelvin, followed by deep laser cooling to achieve higher phase-space densities and longer coherence time \cite{Cheuk2018, Caldwell2019, Caldwell2020} taking advantage of effective trapping in conservative potentials \cite{McCarron2018,Williams2018,Anderegg2018} and grey molasses technique \cite{Ding2020}. Specifically, an optical tweezer array of CaF molecules and further studies of cold collisions of two single molecules have been reported recently \cite{Anderegg2019,Cheuk2020}. Moreover, one-dimensional laser cooling of YbF \cite{Lim2018} and polyatomic molecules, such as, SrOH \cite{Kozyryev2017b}, CaOH \cite{Baum2020}, YbOH \cite{Augenbraun2020} and CaOCH$_3$ \cite{Mitra2020}, have also been realized, making the quantum state manipulation and potential applications be feasible. Besides the above mentioned ones, other species including MgF \cite{Xu2016, Yan2022}, BaH \cite{Tarallo2016}, TlF \cite{Norrgard2017}, AlF \cite{Truppe2019}  and AlCl \cite{Daniel2021} have also attracted great interests. 
\begin{figure}[]
    \centering
    \includegraphics[width=0.48\textwidth]{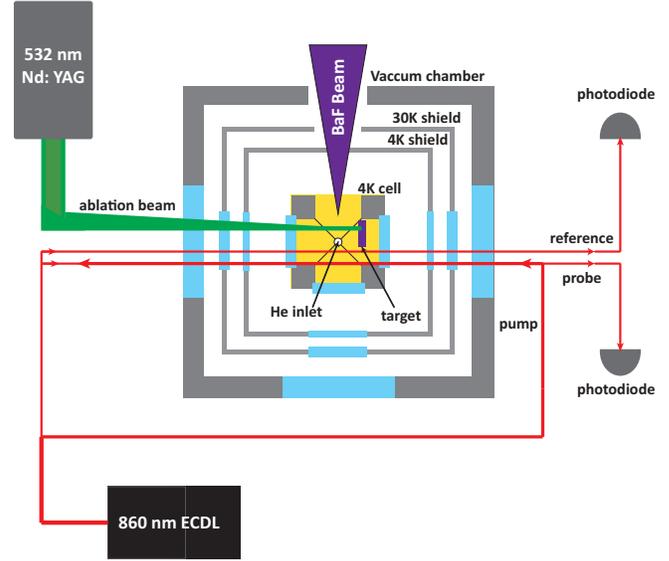}
    \caption{(color online) \label{fig2}
    A schematic illustration of the experimental setup. The three layers from outside to inside are vacuum chamber, $30~{\rm K}$ shield and $4~{\rm K}$ shield, respectively. The  BaF molecules are produced by laser ablation with the Nd: YAG laser in a cell, which is attached to a $4~{\rm K}$ cryogenic head to perform effective buffer gas cooling. The relative position of the BaF$_2$ target and the three laser beams for the spectroscopy detections are also shown. The propagation directions of the probe, pump, and reference beams are indicated by the arrows.
    }
\end{figure}

The BaF molecule which can be potentially laser-cooled \cite{Chen2016} has also received continuous attentions in recent years \cite{Bu2017, Chen2017, Aggarwal2018, Cournol2018, Albrecht2020, Liang2021}. A precise measurement of the relevant transitions is consequently highly required.  Usually, the main cooling transition is chosen as $X^2\Sigma_{1/2}(v=0, N=1,-) \to A^2\Pi_{1/2} (v=0, J=1/2, +)$. The hyperfine structure of the $X^2\Sigma_{1/2}(v=0, N=1,-) $ state has been recognized with high-resolution microwave spectroscopy \cite{Ernst1986}.  But the  hyperfine structure of the excited $A^2\Pi_{1/2} (v=0, J=1/2, +)$ state has not yet been resolved. 

The resolution of the BaF absorption spectroscopy is limited by the Doppler broadening \cite{Bu2017, Albrecht2020}.  One way to get a better resolution is performing the saturated absorption spectrum,  which has been widely used in atomic cases \cite{Hansch1971, Wieman1976, Nakayama1985}.  This has also been applied for YbF  \cite{Skoff2009}, where a resolution of 30 MHz has been reported. Here we extend such a method to the BaF molecule, and further resolve the unknown hyperfine splittings of the excited states. 

The paper is organized as follows. In Sec.\ref{sec3}, a brief description of the experimental setup for the saturated absorption spectroscopy is given. In Sec.\ref{sec4}, we analyze the observed spectra for hyperfine transitions in different branchings and fit them with multi-peak Gaussian function to yield the relevant parameters. Especially, for the main laser-cooling transition $X^2\Sigma_{1/2}(v=0,N=1,-)\leftrightarrow A^2\Pi_{1/2}(v=0,J=1/2,+)$, the fitting shows  the hyperfine splitting of the excited state is about $18~{\rm MHz}$. Finally, we conclude in Sec.~\ref{sec5}.

\section{experimental setup}\label{sec3}

To overcome the Doppler broadening limit and get more knowledge of the laser-cooling relevant transition, we set up a saturated absorption spectroscopy experiment for  BaF molecules. The previous experiment on YbF in Ref.~\cite{Skoff2009} used this technique with a  temperature of 10 K in a liquid helium cryostat. Our experiment here is performed with buffer-gas cooled BaF molecules, under a lower temperature at 4 K produced by a pulse-tube cryostat.

The cryogenic apparatus is built around a pulse tube refrigerator machine (Sumitomo, ${\rm SRP-082B-F70H}$), which has two temperature stages: $30~{\rm K}$ and $4~{\rm K}$. As shown in Fig. \ref{fig2} (a more detailed drawing of the cryogenic apparatus could be found in Ref.~\cite{Bu2017}), the cell in which laser ablation is performed is installed below the 4 K cryogenic head, and surrounded by the 30 K and 4 K shieldings to block the blackbody radiation. The buffer gas  is injected into the cell from the inlet on the back face. The BaF molecules are produced by ablating the BaF$_2$ target with a 532-nm pulsed ND: YAG laser (the repetition rate of the pulse is $1~{\rm Hz}$).

To reduce the systematic errors, we use three laser beams in our experiment, denoted as the pump, probe, and reference beams in Fig.~\ref{fig2}, respectively. These beams are split from an external-cavity semiconductor laser (ECDL), all with {diameters} of $\sim 1~{\rm mm}$. The intensities of the probe and reference beams are set to be nearly equal, at $\sim 20~{\rm mW/cm^{2}}$; while  the pump beam is $\sim 80~ {\rm mW/cm^2}$. We let the probe and the reference beams parallelly enter the cell, while the pump beam counter-propagates from the opposite direction and overlaps with the probe laser in the cell region. The laser frequency is stabilized by a transfer cavity  with the linewidth less than $2~{\rm MHz}$ \cite{Wang2018}. The absorption signals of the probe and the reference beams are monitored by two photodiodes; see Fig.~\ref{fig2}. We scan the laser frequency with a step of $1~{\rm MHz}$, ranging typically several hundreds of ${\rm MHz}$. To obtain a high signal-to-noise ratio, we average the signals by repeating the scanning  hundreds of times.

\section{results and analysis}\label{sec4}


\begin{figure}[htp]
    \centering
    \includegraphics[width=0.47\textwidth]{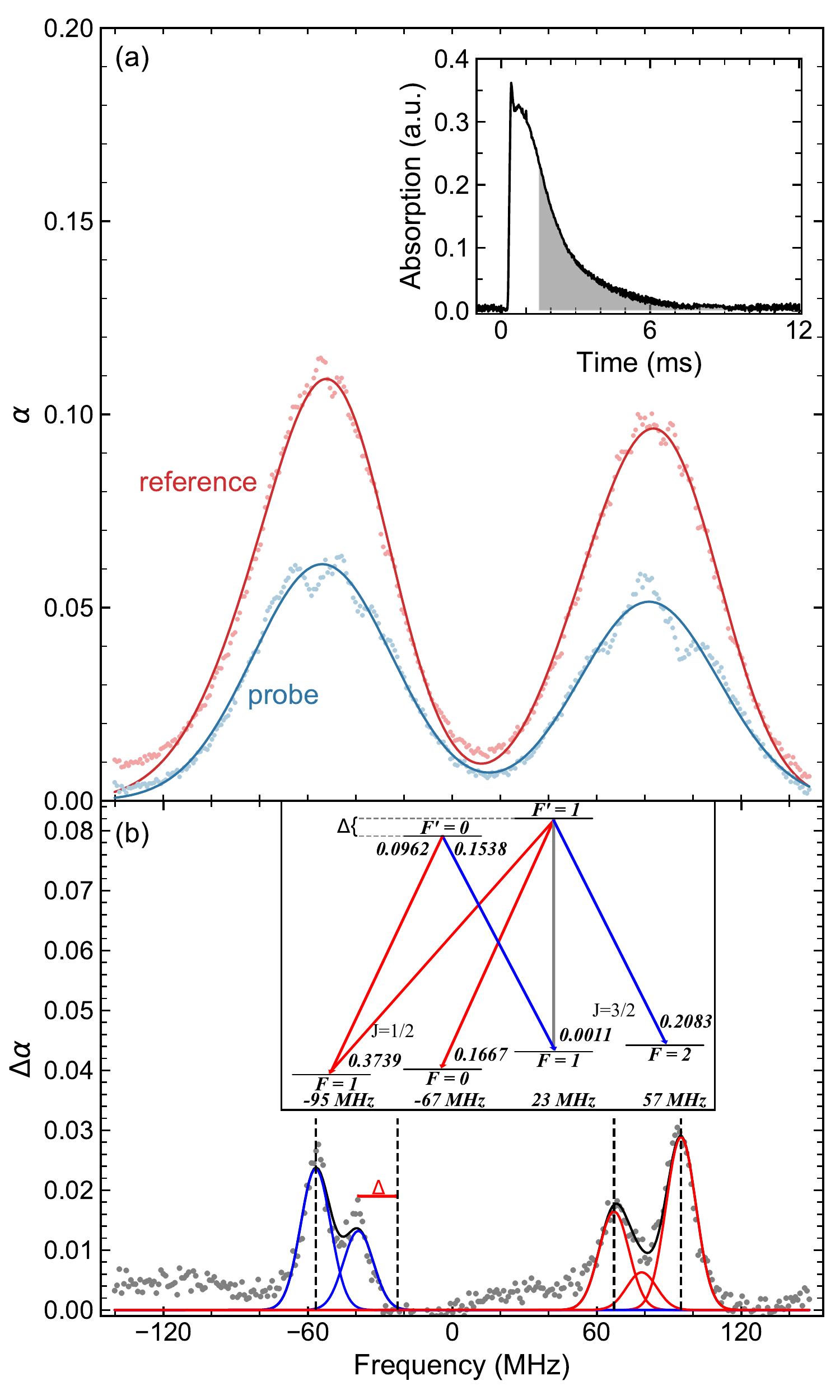}
    \caption{(color online) \label{fig3} 
    The saturated absorption spectroscopy for the main laser cooling transition, $X(v=0,N=1,-)\leftrightarrow A(v=0,J=1/2,+)$, of the BaF molecule.
    (a)  shows the Doppler-broadened absorption spectra (red dots, from the reference beam) and saturated absorption spectra (blue dots, from the probe beam).
 The solid lines are  Gaussian fittings to the two spectra, respectively.
    (b) The Doppler-free saturated absorption spectra. The gray solid line is the fitted curve.  Each peak in our model has been plotted individually,  the color is matched with that in the inset.  The positions of each hyperfine level in the $X(v=0, N=1,-)$ are marked by the vertical black dashed  lines.  The inset shows the corresponding transitions and their branching ratios.
     The transition $F=1(J=3/2) \to F'=1$ is marked by gray, for its branching ratio is negligible.
    }
\end{figure}

{\bf Absorption signal:} A typical time-dependent absorption signal monitored by the photodiode is shown in the inset of Fig.~\ref{fig3}. The ablation laser fires at $t=0~{\rm ms}$, and the absorption signal initially rapidly increases and then exponentially decays. The absorption fraction $A$ is estimated by making a time average of the signal curve from $1.5~{\rm ms}$ to $10~{\rm ms}$ [the shadow region in the inset of Fig.~\ref{fig3}(a)]. We use this time window to avoid the influence  of transient heating of the buffer gas by ablation laser. Then the strength of the absorption can be described by the optical depth $\alpha$ which is related to $A$ as 
\begin{equation}
\alpha=-\ln (1-A)
\end{equation}
 By scanning the laser frequency, we record the $\alpha$'s for both the reference beam and the probe beam as shown in Fig.~\ref{fig3}(a).

{\bf Laser cooling relevant transition:} The absorption spectra for the two beams near-resonant with the main laser cooling transition [$X(v=0,N=1,-)\leftrightarrow A(v=0,J'=1/2,+)$] are shown in Fig.~\ref{fig3}(a). Both of them are Doppler-broadened, and only the fine structure of the ground state can be resolved (the two peaks), similar to our previous observation \cite{Bu2017}. The difference lies in that the spectra from the probe beam show saturated absorption features, i.e., the hole-burning phenomena at the positions of each hyperfine transition. We first rescale the spectra from the probe beam, and then make a subtraction with the two spectra sets to get the Doppler-free saturated absorption spectra in Fig.~\ref{fig3}(b), where the hyperfine sublevels are resolved.

The width of the Doppler-broadened spectra of the reference beam in Fig.~\ref{fig3}(a) reflects the translational temperatures of the buffer-gas-cooled molecules. We fit the spectra with a four-peak Gaussian function [the peak positions are fixed with the values in Fig.~\ref{fig3}(b)]. The averaged full width at half maximum (FWHM) is about $54(4)~{\rm MHz}$, which corresponds to a temperature of $\sim 6.5~{\rm K}$. It is higher than the cell temperature at $4~{\rm K}$, which might be caused by the initial heating of the buffer gas when the ablation laser fires. A similar phenomenon also has been observed in other buffer gas experiments \cite{Barry2011}.

Next, let us analyze the saturated absorption spectra in Fig.~\ref{fig3}(b).   According to the selection rules,  there are six allowed transitions in our sub-doppler spectrum, and the branch ratios are taken from \cite{Chen2019, Albrecht2020}, as shown in Table \ref{table2}. Here, no bias magnetic field is applied in the cell (the earth's magnetic field may exist, but small), so we make an assumption that the Zeeman substates are degenerate. With this, the normalized branching ratios are shown in Fig.~\ref{fig3}(b).
\begin{table}[h]
\caption{\label{table2}
Calculated hyperfine branching ratios for decays from $\vert A,J'=1/2,+\rangle$ state to $\vert X, N=1,-\rangle$ state, also see  \cite{Chen2019, Albrecht2020}.}
\begin{ruledtabular}
\begin{tabular}{ccc|c|ccc}
~ & ~ & ~& ~$F'=0~$ & ~ & $F'=1$ & ~ \\
$J$ & $F$& $m_F$ &$m'_F=0$ ~& $m'_F=-1$ & $m'_F=0$ & $m'_F=1$ \\
\hline
1/2 & 0 & 0 & 0 & 2/9 & 2/9 & 2/9 \\
\hline
~ & ~ & $-1$ & 0.1282 & 0.2493 & 0.2493 & 0 \\
1/2 & 1 & 0 & 0.1282 & 0.2493 & 0 & 0.2493 \\
~ & ~ & 1 & 0.1282 & 0 & 0.2493 & 0.2493 \\
\hline
~ & ~ & -1 & 0.2051 & 0.0007 & 0.0007 & 0 \\
3/2 & 1 & 0 & 0.2051 & 0.0007 & 0 & 0.0007 \\
~ & ~ & 1 & 0.2051 & 0 & 0.0007 & 0.0007 \\
\hline
~ & ~ & -2 & 0 & 1/6 & 0 & 0 \\
~ & ~ & -1 & 0 & 1/12 & 1/12 & 0 \\
3/2 & 2 & 0 & 0 & 1/36 & 1/9 & 1/36 \\
~ & ~ & 1 & 0 & 0 & 1/12 & 1/12 \\
~ & ~ & 2 & 0 & 0 & 0 & 1/6 
\end{tabular}
\end{ruledtabular}
\end{table}

We set the hyperfine splitting between $F'=1$ and $F'=0$ of the $A(v=0, J'=1/2,+)$ to be $\Delta$, as shown in the inset of Fig.~\ref{fig3}(b). The hyperfine splittings of the $X(v=0,N=1,-)$ { are taken from the high-resolution microwave spectra \cite{Ernst1986}}. 
Besides, we assumed that the width of each Gaussian peak is the same, and the relative height of each peak {is proportional to } its corresponding branch ratio, so the fitting formula can be expressed as:
\begin{equation}
\begin{split}
\eta [&0.3739e^{-[(f-95)/w]^2} + 0.1667e^{-[(f-67)/w]^2} + \\ 
&0.1538e^{-[(f+23+\Delta)/w]^2} + 0.2083e^{-[(f+57)/w]^2} + \\
&0.0962e^{-[(f-95+\Delta)/w]^2} + 0.0011e^{-[(f+23)/w]^2}]
\end{split}
\end{equation}
But because one of the transition ( $|J=3/2, ~F=1\rangle \to |F'=1\rangle$ ) has very small branch ratio, it is actually  more like a five-Gaussian function. The best-fitting parameters are:  $w=11.4(6)$, $\Delta = 17.7(2.1)$. The fitting curve and the experimental data agree quite well as shown in Fig.~\ref{fig3}(b).

The parameter $w=11.4(6)$ corresponds to an FWHM of $19(1) {\rm MHz}$ for the transitions observed in our sub-Doppler spectrum. There are several potential broadening mechanisms in our experiment that contribute to this line-width. The main one is the power broadening from the strong pump beam, which can be estimated by the formula, $\Gamma \sqrt{1+s_p}$, with the $A^2\Pi_{1/2}$ state linewidth $\Gamma/2\pi=2.84~{\rm MHz}$ and $s_p$ the saturation parameter. In our experiment, $s_p\approx 45$ leads to a broadening of $\sim 18~{\rm MHz}$, in agreement with our observation. The collisional broadening also plays a role, although less important, in our experiment. A He flow-rate of 2 SCCM corresponds to a density of $1.5\times 10^{15}~{\rm cm^{-3}}$ in the cell, and the collisional cross-section has been measured at an order of $10^{-14}~{\rm cm^2}$ \cite{Bu2017, Albrecht2020}. With these, we estimate the collisional broadening to be $\sim 6~{\rm MHz}$. Other negligible broadening mechanisms include the residual Doppler-broadening due to imperfect alignment of the probe and the pump beams, Zeeman broadening from the Earth's magnetic field (less than $1~{\rm MHz}$), and the transit time broadening.

The parameter $\Delta = 17.7(2.1)$ gives the hyperfine splitting of the the $A(v=0, J'=1/2,+)$ state. The key reason we can resolve such a small splitting lies in that the transition from $|J=3/2, ~F=1\rangle \to |F'=1\rangle$ is almost forbidden, making the peak of $|J=3/2, ~F=1\rangle \to |F'=0\rangle$ well isolated, as shown in Fig. \ref{fig3} (b). Unlike other monofluorides, such as CaF and SrF, the upper-state hyperfine splittings of which are only a few MHz, this value of BaF is relatively large.  Such a large splitting definitely affects the laser cooling efficiency and should be carefully considered in the preparation of the relevant laser frequencies and cooling scheme. The cooling  laser  used in our BaF Doppler cooling experiment has already considered such splitting \cite{Zhang2022}.

\begin{figure*}
    \centering
    \includegraphics[width=0.95\textwidth]{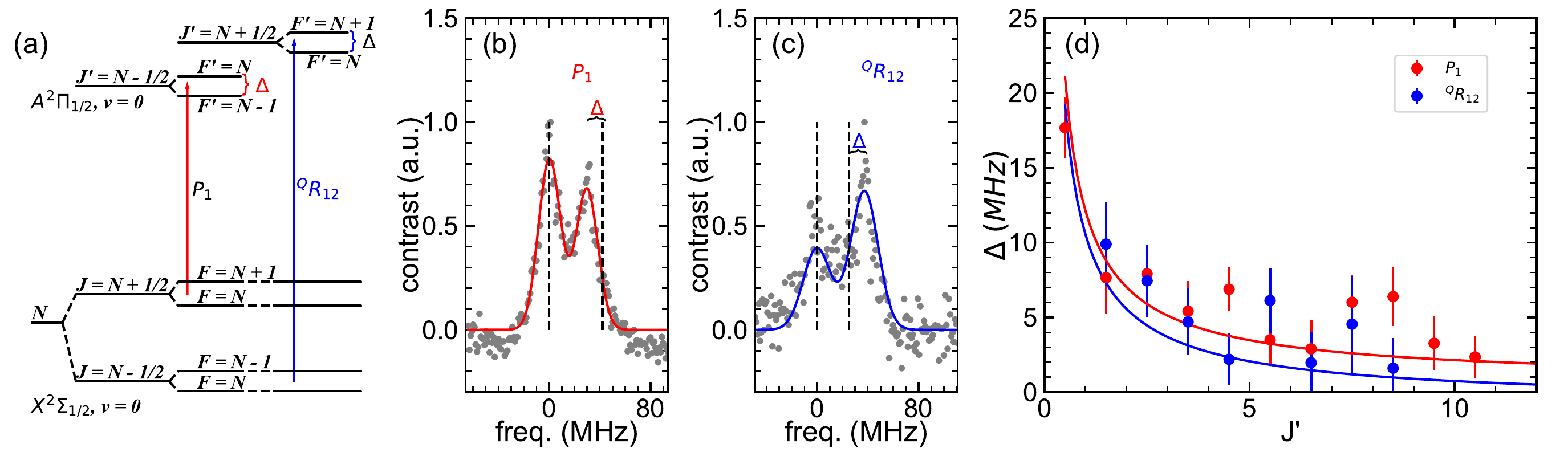}
    \caption{(color online) \label{fig4}
    The hyperfine splitting $\Delta$ of states in $A^2\Pi_{1/2}(v=0)$ measurement: (a) the schematic illustration for the $P_1$ and $^QR_{12}$ branch transitions, which we used to measure the splitting $\Delta$. (b) the typical data measured by the $P_1$ branch [$X(v=0,J=N+1/2)\leftrightarrow A(v=0,J'=N-1/2)$], the black dashed lines marks the positions of the ground states. (c) the typical data measured by $^QR_{12}$ branch [$X(v=0,J=N-1/2)\leftrightarrow A(v=0,J'=N+1/2)$], the positions of the ground states is also marked by the black dashed lines. (d) Distribution of the measured $\Delta$ in different $J'$ state, the red and blue dots are obtained from $P_1(N)$ and $^QR_{12}(N)$ branches, respectively. 
{The solid lines show the  fitting curves accordingly.}
}
\end{figure*}

{\bf Other transitions with high $J'$:} We further extend our measurement to the hyperfine splittings $\Delta$ of states with a higher value of $J'$ in $A^2\Pi_{1/2}(v=0)$. We measure the $P_1$ and $^QR_{12}$ branch transitions, then use a similar fitting method as described above, we could obtain the $\Delta$'s for each $J'$ state. The transitions of these two branches are illustrated in Fig.~\ref{fig4}(a). According to the illustration and the selection rules, there are three allowed hyperfine transitions for a $P_1$ transition ($J \to J'=J-1$). We consequently set the fitting function as a three-Gaussian model
\begin{equation}
\begin{split}
\eta\{r_{N+1, N}&e^{-(f/w)^2} + r_{N, N}e^{-[(f-\Delta_0)/w]^2} +\\
& r_{N, N-1}e^{-[(f-\Delta_0+\Delta)/w]^2}\}
\end{split}
\end{equation}
Similarly, the $^QR_{12}$ branch spectrum should be fitted with the formula:
\begin{equation}
\begin{split}
\eta\{r_{N-1, N}&e^{-(f/w)^2} + r_{N, N}e^{-[(f-\Delta_0)/w]^2} +\\
& r_{N, N+1}e^{-[(f-\Delta_0-\Delta)/w]^2}\}
\end{split}
\end{equation}
where $r_{ij}$ is the branching ratio of the transition from $F = i$ to $F' = j$, and $\Delta_0$ is the relative position of the hyperfine level of the ground state. We set the state $F = N + 1$ and $F = N - 1$ to be at the zero position for $P_1$ and $^QR_{12}$ branches, respectively. The $r_{ij}$ and $f$  can be calculated by using the molecular constants of BaF measured in \cite{Ryzlewicz1980, Ernst1986}. The other parameters $\eta, w$, and $\Delta$ are fitting parameters, and here we are only concerned about $\Delta$.

Fig.~\ref{fig4}(b) and Fig.~\ref{fig4}(c) show the typical data measured in our experiment, corresponding to one transition line for $P_1$ branch and $^QR_{12}$ branch, respectively. Note that we only observed two transitions in the experiment for both branches, although we have mentioned above that there should be three transitions. This is because the branching ratio of $r_{N,N}$ in both $P_1$  and $Q^R_{12}$ transitions is negligible. For sufficiently large $J'$, there is a quasi-selection rule $F'- F = J'- J$, which makes the transitions reduce to two. For the two transitions observed in the spectra, due to the hyperfine splitting $\Delta$ of the excited state, the distance between the two observed transition peaks is shifted from the interval of the ground states by $\Delta$. As shown in Fig.~\ref{fig4}(b) and Fig.~\ref{fig4}(c), the distance of two peaks in $P_1$ branch is smaller than the interval of the ground states by $\Delta$, while oppositely, for $^QR{12}$ branch, it becomes larger by $\Delta$.

In Fig.~\ref{fig4}(d) we plot the measured $\Delta$.  The hyperfine splittering of BaF of $A^2\Pi_{1/2}$ is \cite{Sauer1999}
\begin{equation}\label{fitEq}
\Delta = \left(\frac{A_\parallel}{2J'+1}+\eta A_\perp\right)\frac{(2J'+1)^2}{8J'(J'+1)}
\end{equation}
where $\eta=-1$ for $J'=N-1/2$ states and $\eta=+1$ for $J'=N+1/2$ states.
The best-fit  gives $A_\parallel = 60.4(5.4)~$MHz and $A_\perp=-1.4(1.0)$ MHz,  and the fitting curves are shown in Fig.~\ref{fig4}(d),

Finally, let us discuss the origin of the uncertainty, i.e., the error bars in Fig.~\ref{fig4}(d). The main source is the nonlinear effect of the PZT of the Fabry-Perot cavity. As described in Ref.~\cite{Wang2018, WBu2016}, we developed a transfer cavity lock system for BaF
laser-cooling experiment. We determine the frequency of our laser from its absorption position in the Fabry-Perot cavity. When we transform the position to frequency, we assumed that their relationship is linear. However, there is a slight nonlinear effect for the PZT of the Fabry-Perot cavity, which makes the linear correspondence not as accurate as expected. For the measurements of $P_1$ and $^QR_{12}$ branches, the interval between two peaks is about 30MHz. By using the results from Ref.~\cite{WBu2016},  we declare the error due to the nonlinear effect of PZT is about 1MHz.
Another systematic error is the linewidth of our locked laser. We have measured it in our previous work \cite{Wang2018}, which is less than 1MHz. Therefore we estimate that the error caused by the laser linewidth is 1MHz.
The fitting error is 68 \% confidence interval provided by the fit function and already included in Fig.~\ref{fig4}(d). Errors caused by other mechanisms should be negligible in our experiment.

~\\

\section{conclusion}\label{sec5}
To conclude, we have employed the Doppler-free saturated absorption scheme to measure the hyperfine splittings of each $J'$ in the excited $A(v=0)$ state of the BaF molecule. The information of the hyperfine splittings provides fundamental foundations for future laser cooling experiments on BaF molecules. And the knowledge of the hyperfine splittings of the higher $J'$ ($J' > 1/2$) states is useful to gain more information about the molecular structures for the BaF molecule.

\begin{acknowledgments}
We thank Prof. Eric Hessels for useful discussions, and acknowledge the support from the National Key Research and Development Program of China under Grant No. 2018YFA0307200, the National Natural Science Foundation of China under Grant Nos. U21A20437 and 12074337, Natural Science Foundation of Zhejiang Province under Grant No. LR21A040002, Zhejiang Province Plan for Science and technology No. 2020C01019, and the Fundamental Research Funds for the Central Universities under No.2021FZZX001-02.
\end{acknowledgments}

\bibliographystyle{apsrev4-1}
\bibliography{saturation}

\end{document}